\documentclass[aps,prd,preprint,amsmath,superscriptaddress]{revtex4}

\usepackage{graphicx}

\begin{document}
\title{ 
Gamma rays from dark matter annihilation in the Draco and
observability at ARGO
}

\author{Xiao-Jun Bi}
\email{bixj@mail.ihep.ac.cn}
\author{Hong-Bo Hu}
\affiliation{Key laboratory of particle astrophysics, IHEP}
\author{Xinmin Zhang}
\affiliation{Theoretical Physics Division, IHEP\\
Chinese Academy of Sciences, Beijing 100049, P. R. China}

\begin{abstract}
The CACTUS experiment recently observed a gamma ray excess 
above 50 GeV from the direction of the Draco dwarf spheroidal galaxy.
Considering that Draco is dark matter dominated the
gamma rays may be generated through dark matter annihilation
in the Draco halo. In the framework of the minimal supersymmetric
extension of the standard model we explore the parameter space
to account for the gamma ray signals at CACTUS. We find that the
neutralino mass is constrained to be approximately in the range
between 100 GeV $\sim$ 400 GeV and a sharp central cuspy of the dark
halo profile in Draco is necessary to explain the CACTUS results. 
We then discuss further constraints on the supersymmetric
parameter space by observations at the ground based ARGO detector. 
It is found that the parameter space can be strongly constrained by 
ARGO if no excess from Draco is observed above 100 GeV .
\end{abstract}

\maketitle

\section{ introduction }

The existence of cosmological dark matter has been established by
various astronomical observations.
However, the evidences come mainly from the gravitational effects 
of the dark matter component. The nature of dark 
matter  remains elusive and keeps one of the most outstanding 
puzzles in particle physics and cosmology \cite{review}.
The primordial nucleosynthesis and
cosmic microwave background measurements constrain the baryon component
and most dark matter component should be non-baryonic.
The development in understanding the large scale structure formation  
requires the dark matter be cold. From the theoretical
considerations the favored candidate for cold dark matter (CDM) 
seems to be the weakly interacting massive particles (WIMPs) 
\cite{review}.

The WIMPs can be detected indirectly by observing the annihilation 
products, such as gamma rays, neutrinos, anti-protons
and positrons. Exploring the anomalous results from the
cosmic ray experiments is one viable way to identify the dark matter.
Since the annihilation rate is proportional to the square of the
dark matter density, the ideal sites for dark matter detection 
should have high dark matter density. The galactic center is 
believed to be a promising source of dark matter annihilation \cite{berg}.
However, the existence of the central supermassive black hole and
the supernova remnant Sgr A$^*$ contaminates the dark matter
signals heavily . Alternative sites, such as the substructures of
the Milky Way or the dark matter dominated dwarf spheroidal galaxies (dSph),
have been studied in the Refs\cite{baltz, pieri, evans, bi, kou}.

Recently, the CACTUS gamma-ray experiment reported an excess of gamma
rays from the direction of Draco, a nearby dSph \cite{cactus}.
Since Draco is dark matter dominated and no other gamma ray sources
are expected to be hosted \cite{young, tyler} the excess has been attributed
to the annihilation of dark matter in the Draco halo \cite{hooper, profumo}.
The results are still preliminary and, if confirmed, will have important
implications on the nature of dark matter and the density profile
of Draco.  Additional observations of the signal
by other experiments is therefore very important. 
The GLAST \cite{glast}, a satellite-based experiment, 
and the MAGIC \cite{magic}, a ground-based
Atmospheric \v{C}erenkov Telescope (ACT), have been considered to
check the CACTUS results \cite{hooper, profumo}. In the present
work, we will discuss a possibility of detecting or
constraining the gamma rays observed by CACTUS at ARGO \cite{argo}, 
a ground-based extensive air shower (EAS) detector.

In the next section we will first give the general formula
for dark matter annihilation. Then we will discuss the implications
of CACTUS results on the gamma ray spectrum and fluxes in Sec III.
The sensitivity of ARGO is given in Sec. IV and the numerical
results are presented in Sec. V. We conclude in Sec. VI.

\section{ gamma rays from dark matter annihilation }

The annihilation of two WIMPs can produce the continuous spectrum of
gamma rays arising mainly
in the decays of the neutral pions produced in the fragmentation processes
initiated by the tree level final states. The fragmentation and 
decay processes can be simulated with the Pythia package\cite{pythia}. 

The annihilation rate in unit time and unit volume is given by
\begin{equation}
R=\langle \sigma v \rangle n^2/2=\frac{\langle\sigma v\rangle\rho^2}{2m^2}
\end{equation}
where $\sigma$ and $v$ are the annihilation cross section and
the relative velocity of the two dark matter particles respectively,
$n$ and $\rho$ are the number and mass densities of dark matter
and $m$ is its mass, the factor $2$ in the denominator
arises due to the identical initial particles.
We note that the annihilation rate is proportional to the square
of the dark matter density and therefore,
a high density region can greatly enhance
the annihilation fluxes.
                                                                                
The gamma ray flux from the Draco halo is therefore given by
\begin{equation}
\label{flux}
\Phi_\gamma ( E)=\phi^\gamma(E)\frac{\langle\sigma v\rangle}{2m^2}
 \frac{ \int dV \rho^2}{4\pi D^2} = \frac{\phi^\gamma(E)}{4\pi }
\frac{\langle\sigma v\rangle}{2m^2}\times \frac{1}{D^2}
\int_{\Delta \Omega}d\Omega
\int 4\pi r^2 dr\rho^2(r)
\end{equation}
where the halo profile is assumed approximately spherically symmetric
with the density profile $\rho(r)$,
$D=75.8\pm0.7\pm5.4$ kpc is the distance to Draco \cite{cepheid},
$\phi^\gamma(E) $ is the differential flux at energy $E$ 
in a single annihilation in unit of 1 gamma GeV$^{-1}$.
$\Delta \Omega$ represents the angular resolution of the detector.

The density profile $\rho(r)$ of Draco is constrained by observations.
A recent analysis shows that both a cored and a cuspy profile,
such as the NFW profile \cite{nfw}, are
consistent with the observational data and the results of N-body
simulation \cite{mash}. The `astrophysical factor' in Eq. (\ref{flux})
defined as 
\begin{equation}
\Phi_{astro} = \frac{1}{D^2}\int_{\Delta \Omega}d\Omega
\int 4\pi r^2 dr\rho^2(r)\ ,
\end{equation}
which is determined by the astrophysical quantities solely, is
severely constrained by observational data. It is found that 
$\Phi_{astro}$ varies by a factor of approximately only
$200$, i.e., 
$\Phi_{astro} \cong (3.2\times 10^{-4} \sim 6.4 \times 10^{-2})$ GeV$^2$
cm$^{-6}$ kpc sr following Ref. \cite{mash}.

The other part in Eq. (\ref{flux})
is determined by particle physics which defines the nature of
dark matter. We will calculate the `particle factor' in the
framework of the minimal supersymmetric standard model (MSSM).
The MSSM is the most attractive model beyond the standard model
of particle physics. In the R-parity conserved MSSM, the lightest
supersymmetric particle, the lightest neutralino, provides a
natural candidate for WIMP. The MSSM is well defined by a 
set of
free parameters, which lead to the uncertainties in predicting
the gamma ray flux from the particle physics.
Once the particle factor is determined and combine with the
astrophysical factor given above, we can give the
predicted gamma ray flux from Draco.

\section{ the CACTUS experiment }

CACTUS is a ground based Air \v{C}herenkov Telescope (ACT) 
located at Solar Two near Barstow, California.
CACTUS utilizes a set of 144 heliostats, each 42 $m^2$, 
to form a composite mirror with a total effective area of about 6,000 $m^2$.
The threshold energy for gamma rays at CACTUS is about 50 GeV and
the effective area for $\gtrsim 200 GeV$ gamma rays reaches 
about 50,000 $m^2$.
 
Within the angular region of about $1^\circ$ centered around 
the direction of Draco, CACTUS has recently observed an excess 
of approximately 30,000 photons for 7 hours observation 
above the average background outside Draco\cite{cactus}.
The threshold energy of the photons is about 50 GeV. There
is no significant excess observed if the cutoff energy is improved
to about $150 GeV$. Although the results are still preliminary, however,
if confirmed, the implications on dark matter are significant.
It is interesting to consider the implications of the CACTUS 
experimental results
seriously due to our completely ignorance of the nature of dark matter.
In this section we will study the implications for the gamma ray
spectrum and flux from the CACTUS results.

The gamma events are given by
\begin{equation}
\label{ngamma}
N_\gamma^{\rm observed}
= \epsilon_{\Delta \Omega}\int_{E_{th},\Delta \Omega}^{m_\chi} 
A_{eff}(E) \Phi(E) dE d\Omega dT\ \ ,
\end{equation}
where $\epsilon_{\Delta \Omega}=0.68$ is the fraction of signal
events within the angular resolution of the instrument and the
integration is for the energies above the threshold energy $E_{th}$ and
below the mass of neutralino, $m_\chi$, within
the angular resolution of the instrument $\Delta \Omega$ and for the
observational time. The effective
area $A_{eff}$ is a function of energy and
$\Phi(E) = \phi_0 \frac{dN_\gamma}{dE}$ is the flux of 
$\gamma$-rays from DM annihilation with $\phi_0$ the intensity 
normalization and $\frac{dN_\gamma}{dE}$ the shape of the spectrum.
The effective area of CACTUS, which is energy dependent, 
can be parametrized as 
\begin{equation}
\label{aeff}
A_{\rm{eff}} \approx 47,000 \, \rm{m}^2 \, \left[ 1-e^{-0.014 \,(E_{\gamma}-39.6\, 
\rm{GeV})}\right] + 11.9 \times \, E_{\gamma}(GeV)\ ,
\end{equation}
due to the simulation results \cite{cactus}.

\begin{figure}
\includegraphics[scale=1]{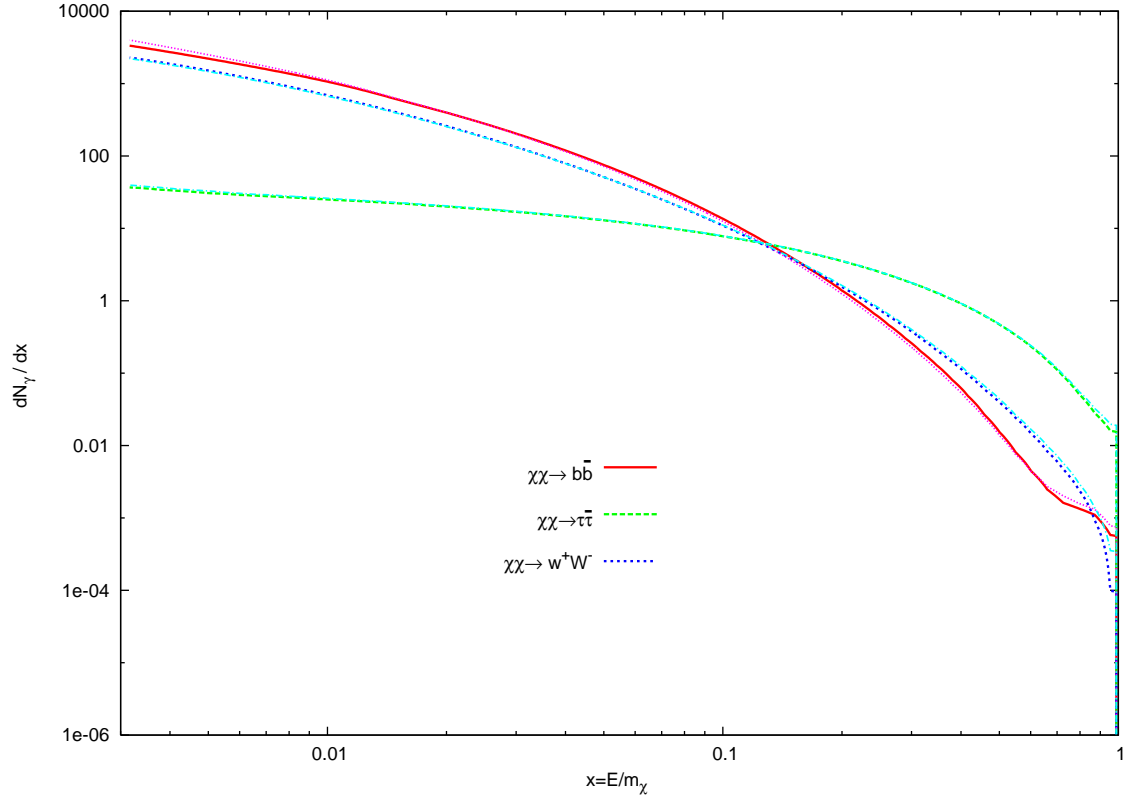}
\caption{\label{spec}
The spectrum of gamma rays from neutralino annihilation, 
$\frac{dN_\gamma}{dx}$ with $x=E_\gamma/m_\chi$, for the
final state of $W^+W^-$, $b\bar{b}$ and $\tau\bar{\tau}$. 
$m_\chi=100, 500 GeV$ has been taken which gives almost 
identical spectrum $\frac{dN_\gamma}{dx}$ for each final state. }
\end{figure}

From Eq. (\ref{ngamma}), we can see that in order to obtain the gamma
ray flux from the observed event number
we have to assume the gamma ray spectrum first.
The spectrum of the gamma rays through neutralino annihilation
depends on the final states into which
the neutralinos have annihilated. In Fig. \ref{spec} we show the
spectrum of gamma rays for the final states of gauge bosons 
$\chi\chi\to W^+W^-$ 
and for the final states of $\chi\chi\to b\bar{b}$ and 
$\chi\chi\to\tau\bar{\tau}$, which represent the two extreme cases that 
the annihilated gamma rays have soft and hard spectra respectively.
In the figure we have plotted the spectrum for $m_\chi=100, 500 GeV$
respectively.  We find the spectrum, expressed
as a function of a dimensionless quantity $x=E_\gamma/m_\chi$,
is not sensitive to the mass of the neutralino, $m_\chi$.

The integrated gamma ray flux above the threshold energy of $50 GeV$
is given by
\begin{equation}
I_\gamma(> 50 GeV) = \int_{50 GeV}^{m_\chi} \Phi(E)dE
=\int_{50 GeV}^{m_\chi} \frac{dN_\gamma}{dE} dE \cdot
\frac{N_\gamma^{\rm observed}}{\epsilon_{\Delta \Omega}\int_{50 GeV}^{m_\chi }
A_{eff}(E) \frac{dN\gamma}{dE} dE \cdot T}\ \ ,
\end{equation}
where we have assumed that the effective area has no zenith angle
dependence within the angular resolution.
From this equation we know the softer the spectrum the greater 
the gamma ray flux since $A_{eff}$ is small at low energies.
For soft spectrum, taking $m_\chi=100 GeV$ and $b\bar{b}$ final states, 
we get
$I_\gamma(> 50 GeV) = 1.7\times 10^{-8} cm^{-2} s^{-1}$, while
for the hard spectrum, taking $m_\chi=300 GeV$ and $\tau\bar{\tau}$ 
final states, we get
$I_\gamma(> 50 GeV) = 7.3 \times 10^{-9} cm^{-2} s^{-1}$.
This spectrum is taken in order not to give too much excess above
150 GeV.
Concerning the uncertainties from the noise rejection procedures,
the misidentification of the electronic and hadronic primary events
and that the angular region of CACTUS is larger than that of Draco,
the observed excess may be greatly larger than the real {\em signal}
of dark matter annihilation. Therefore in our theoretical calculation 
we make an assumption that the uncertainty of the gamma ray flux 
is larger than the current CACTUS data by relaxing the lower bound 
by an order of magnitude. We finally get the gamma ray flux 
from Draco which is approximately in the range of 
\begin{equation}
\label{i50}
7.3 \times 10^{-10} < I_\gamma(> 50 GeV)< 
1.7\times 10^{-8} cm^{-2} s^{-1}\ .
\end{equation}  

Since there is no significant excess observed above 150 GeV the gamma ray 
spectrum is further constrained. We assume that the events above 150 GeV
do not exceed
the Possion fluctuation of the background, which includes the misidentification
of hadronic cosmic rays as gamma signals, the electronic comic ray events 
and the galactic diffuse gamma rays. We have adopted the expressions as
\begin{equation}
\phi_h(E) =1.49 E^{-2.74} cm^{-2}s^{-1}sr^{-1}GeV^{-1}
\end{equation}
for the hadronic contribution \cite{gaisser},
\begin{equation}
\phi_e(E) =6.9\times 10^{-2} E^{-3.3} cm^{-2}s^{-1}sr^{-1}GeV^{-1}
\end{equation}
for the electronic contribution \cite{longair},
\begin{equation}
\phi_{\text{galac}-\gamma}(E) =8.56 \times 10^{-6} E^{-2.7} 
cm^{-2}s^{-1}sr^{-1}GeV^{-1}
\end{equation}
for the Galactic $\gamma$-ray emission at the direction of Draco
($l=86.4^\circ$, $b=34.7^\circ$), extrapolated from the EGRET
data at low energies\cite{gal}.

In principle the gamma ray flux above 150 GeV also depends on
the spectrum of the gamma ray. However, due to Eq. (\ref{aeff})
the effective area above 150 GeV is not so sensitive to energy differing 
from that
at energies below 100 GeV. Considering the large systematic uncertainties
and the possible problems in the noise reduction procedure
we approximate the  effective area above 150 GeV as 50,000 m$^2$, being
a constant. Then we get a conservative upper limit of $I_\gamma(> 150 GeV)$.
Assuming that about 90 percent of the hadronic comic ray background can be
rejected within the angular region due to the different shape of
the \v{C}herenkov wavefront induced by electronic and hadronic showers, 
we get that 
\begin{equation}
\label{i150}
I_\gamma(> 150 GeV) \lesssim 3.\times 10^{-11} cm^{-2}s^{-1}\ .
\end{equation}

In the next sections we will explore the supersymmetric (SUSY) parameter
space to account for the gamma excess observed at CACTUS 
taking into account the constraints given by
the Eqs. (\ref{i50}) and (\ref{i150}).

\section{ sensitivity of ARGO }

The ARGO-YBJ experiment, locates at YangBaJing 
(90.522$^\circ$ east, 30.102$^\circ$ north, 4300m a.s.l) 
in Tibet, China, is a ground-based telescope optimized 
for the detection of small size air showers. The energy 
threshold of the detector is designed to be about 100GeV. 
The detector consists of a single layer of RPCs floored 
in a carpet structure covering an area of $\sim 10^4 m^2$. 
The detector is under construction and the central carpet will 
be completed early in 2006 and put in stable 
data taking soon after.

The performances of the detector have been studied by 
means of Monte Carlo simulations \cite{cui}. Defined as 
a product of the sampling area and the trigger efficiency, 
the effective area characterizes the power of the detector in 
recording the number of events for a given energy and time 
interval from a given direction. For both primary $\gamma$ and 
hadrons with energy near the threshold, the effective area 
can be approximately parameterized as $A_{\rm eff}\approx A_{100} E^{2.4}$, 
when the trigger condition is set to be larger than or equal to 20 
fired pads, where $A_{100} \sim 100 m^2$ is the effective area 
for primary $\gamma$ ray events at the threshold energy of about 
100 GeV \cite{cui}. Above the threshold energy the effective area increases 
rapidly and reaches about 10,000 $m^2$ for TeV gamma rays.
At the same time, simulation also shows that at low
energies the protons have lower trigger efficiency than photons.
The effective area for protons near the threshold energy is about
one order of magnitude smaller than that of gamma,
leading to a great suppression of the background.

The Draco dSph is within the field of view of the ARGO detector with 
the closest zenith angle to be $\sim 27^\circ$. Following up
observations on the gamma excess seen by CACTUS have been considered 
at GLAST and MAGIC \cite{hooper, profumo}. Ground-based extensive 
air shower (EAS) arrays with low energy threshold, such as ARGO \cite{argo} 
and the next generation all-sky high energy gamma-ray telescope 
HAWC \cite{hawc}, have complementary properties to the satellite 
borne experiments and the ACTs.  They have large effective areas 
and at the same time possess the advantages in large field of view 
and near 100\% duty cycle. However, the EAS arrays usually have a 
poorer hadron-photon identification power. In this work, we will 
discuss how to constrain the gamma ray signal from Draco by the
ARGO experiment.

For this purpose, we focus on the events for the energy 
below $\sim 400 GeV$, since we will see in the next section that 
the CACTUS excess constrains the neutralino mass to be lower than
$\sim 400 GeV$. The number of background events for one year's 
data taking at ARGO is therefore also estimated in this energy range.
To constrain the signal at the $2\sigma$ level
for one year's observation, the flux above 50 GeV from 
Draco is then constrained as
\begin{equation}
I_\gamma( > 50 GeV) = \frac{2\sqrt{N_{bkg}}}{
A_{100}T}\cdot \frac{\int_{50}^{m_\chi}\frac{dN_\gamma}{dE}dE}
{\epsilon_{\Delta\Omega} \int_{100}^{m_\chi}
\left(\frac{E}{100}\right)^{2.4}\frac{dN_\gamma}{dE}dE}\ ,
\end{equation}
where again the zenith angle dependence of the effective area
of the ARGO detector is ignored.



\section{ numerical results }

\begin{figure}
\includegraphics[scale=1]{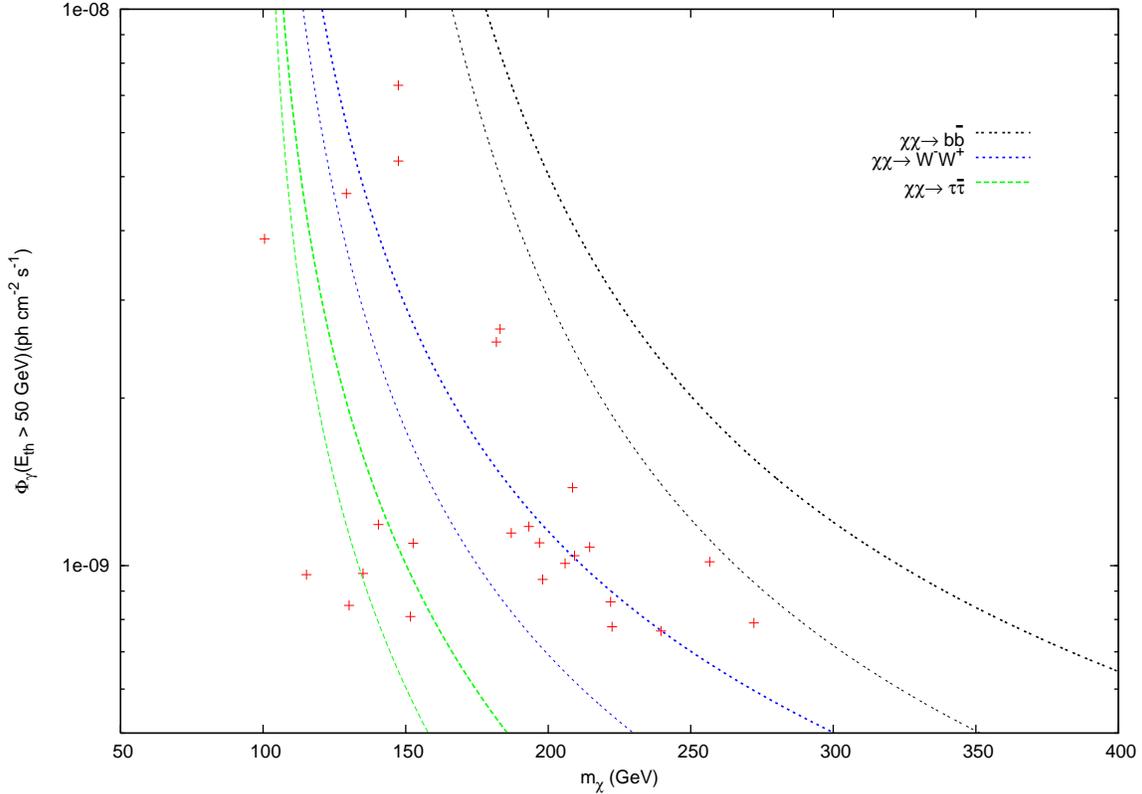}
\caption{\label{10}
The integrated $\gamma$-ray fluxes by neutralino annihilation 
from Draco above the threshold energy of  $50$ GeV
as a function of the neutralino mass.
The fluxes are given within the angular resolution of
$\Delta\Omega =10^{-3}$. Each point in the figure represents a set
of low energy SUSY parameters which survive all the current limits.
A boost factor $10$ relative to the maximal astrophysical factor
derived from \cite{mash} has been assumed.
The lines shows the 2$\sigma$ constraints from the ARGO experiment
assuming a $W^+W^-$, $b\bar{b}$ or $\tau\bar{\tau}$ final state with
or without gamma/hadron discrimination.
}
\end{figure}
                                                                                
\begin{figure}
\includegraphics[scale=1]{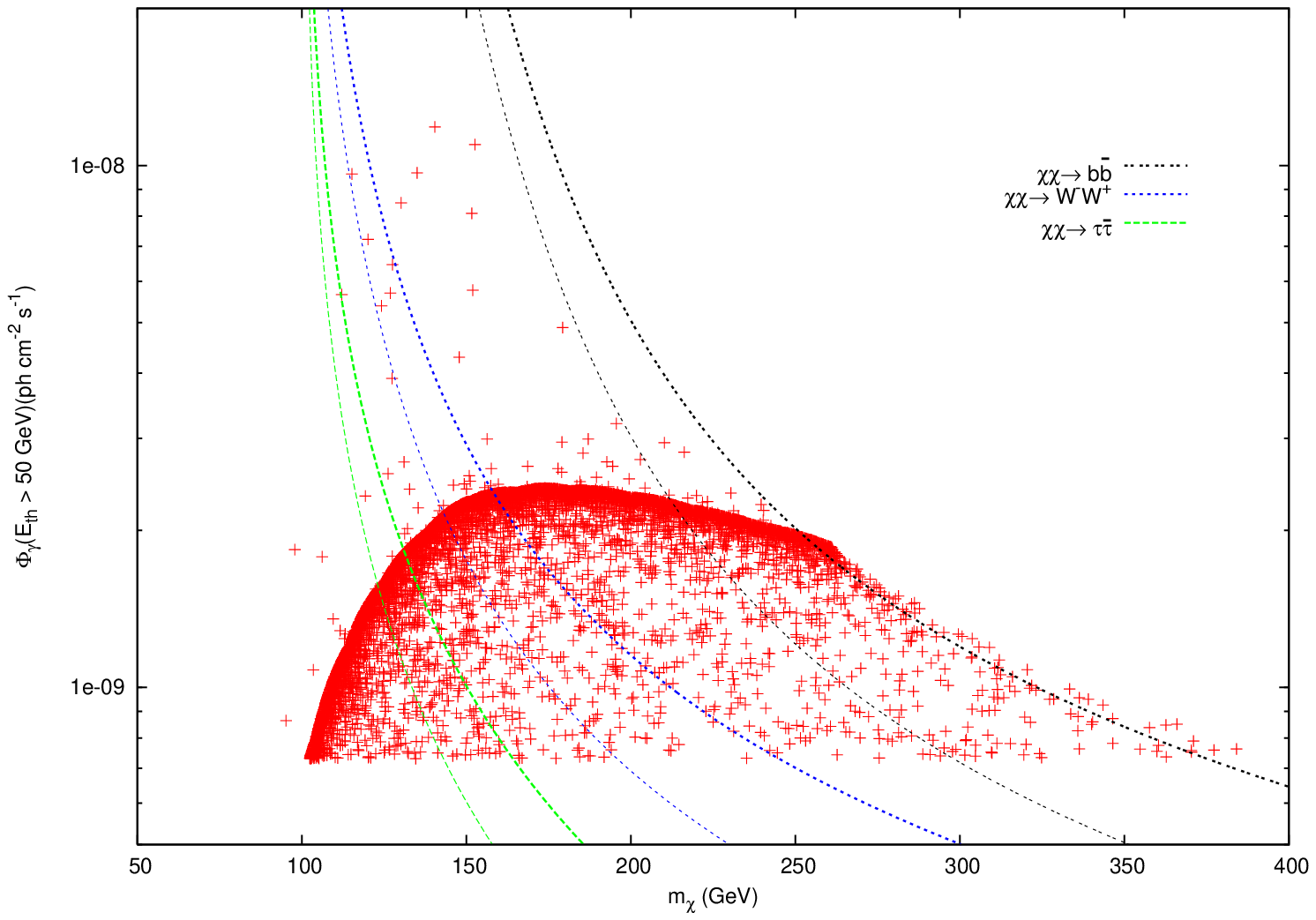}
\caption{\label{100}
Same as Fig. \ref{10} except that a boost factor of $100$
has been assumed.}
\end{figure}
                                                                                
\begin{figure}
\includegraphics[scale=1]{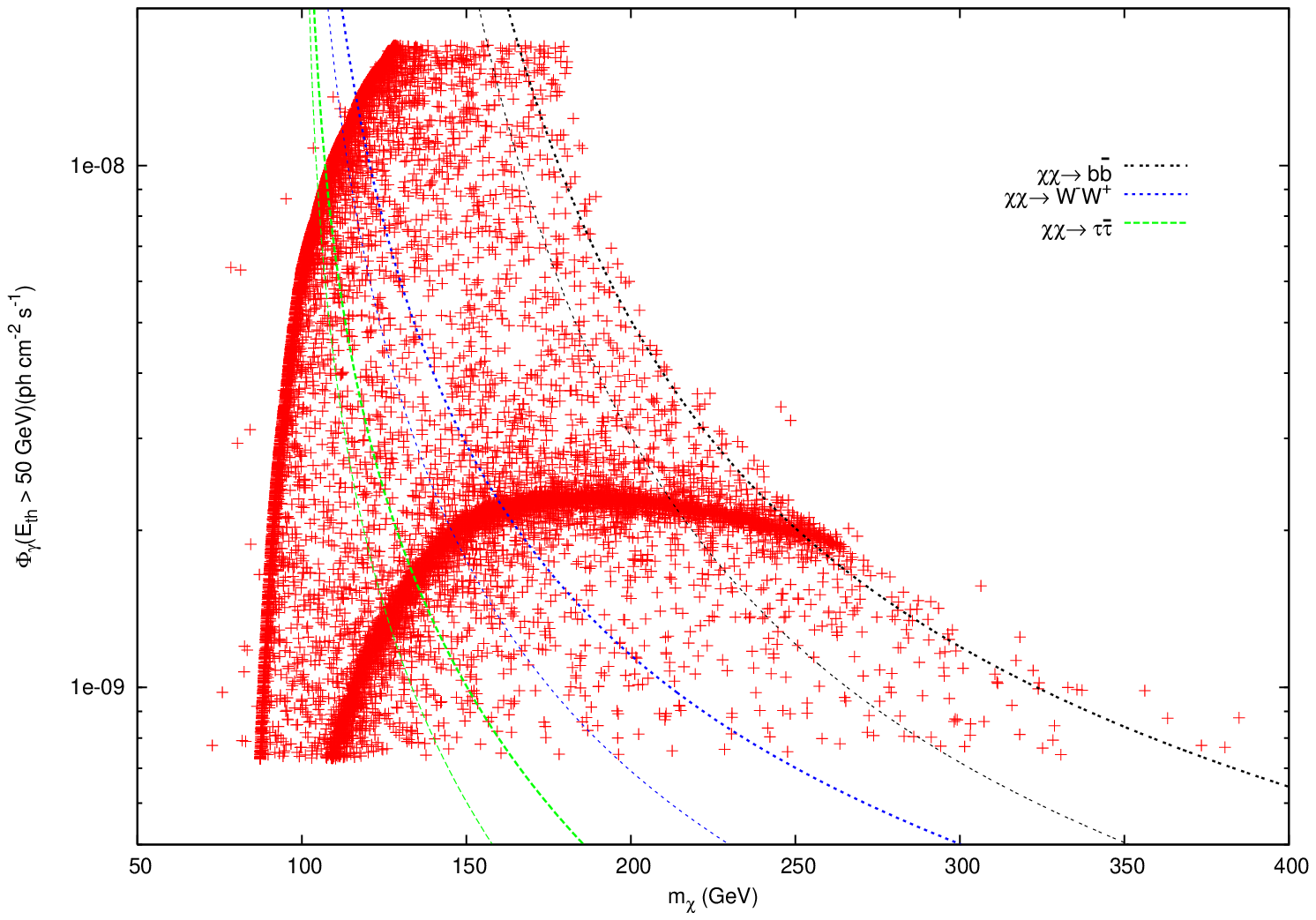}
\caption{\label{1000}
Same as Fig. \ref{10} except that a boost factor of $1000$
has been assumed.}
\end{figure}
                                                                                
In this section we will explore the SUSY parameter space to account
for the CACTUS excess assuming that the excess (or a fraction of the 
excess) is generated by neutralino annihilation in the Draco halo.
The constraint on the parameter space from ARGO is taken into account.

The  R-parity conserved MSSM
is described by more than one hundred parameters describing the soft
supersymmetry breaking. However, for
the processes related with dark matter production and annihilation,
only several parameters are relevant under some simplifying assumptions,
i.e., the higgsino mass parameter $\mu$, the bino mass parameter $M_1$,
 the wino mass parameter $M_2$,
the mass of the CP-odd Higgs boson $m_A$, the ratio of the Higgs
vacuum expectation values $\tan\beta$, the scalar fermion mass parameter
$m_{\tilde{f}}$, the trilinear soft breaking parameter $A_t$
and $A_b$. To determine the low energy spectrum of the SUSY particles
and coupling constants,
the following assumptions have been made: all the sfermions
have common soft-breaking mass parameters $m_{\tilde{f}}$;
all trilinear parameters are zero except those of the third 
family; the gluino and wino have the
mass relation, $M_3=(\alpha_s(M_Z)/\alpha_{em})
\sin^2\theta_W M_2$, coming from the unification
of the gaugino mass at the grand unification scale.
However, to explore more general low energy phenomenological SUSY parameter
space we relax the relationship between $M_1$ and $M_2$ derived from
the grand unification scale.
                                                                                
We perform a numerical random scanning in the 8-dimensional
supersymmetric parameter space using the package DarkSUSY
\cite{darksusy}. The ranges of the parameters are as following:
$50 GeV < |\mu|,\ M_2,\ M_A,\ m_{\tilde{f}} < 5 TeV$,
$1.1 < \tan\beta < 60$, $-3m_{\tilde{q}} < A_t, \ A_b < 3m_{\tilde{q}}$,
$\text{sign}(\mu)=\pm 1$.
The parameter space is constrained by the theoretical consistency
requirements, such as the correct symmetry breaking pattern,
the neutralino being the LSP and so on. The accelerator data
constrains the parameter further
from the spectrum requirement, the invisible Z-boson width,
the branching ratio
of $b\to s\gamma$ adopting the experimental data given by the Particle
Data Group in the year of 2002\cite{pdg}.
                                                                                
Another important constraint comes from cosmology. Combining the recent
observation data on cosmic microwave background, large scale structure,
supernova and data from HST Key Project
the cosmological parameters are determined quite precisely. Especially,
the abundance of the cold dark matter is given by \cite{wmap}
$\Omega_{\text{CDM}}h^2=0.113^{+0.008}_{-0.009}$.
We constrain the SUSY parameter
space by requiring the relic abundance of neutralino
$0 < \Omega_\chi h^2< 0.137$,
where the upper limit corresponds to the 3$\sigma$ upper bound from
the cosmological observations. When the relic abundance of neutralino
is smaller than a minimal value the thermally produced neutralino 
represents a subdominant dark matter component. We assume non-thermal
mechanism to give the correct dark matter relic density\cite{nonther}.
The effect of coannihilation between the fermions is taken into account
when calculating the relic density numerically.
                                                                                
We find that a `boost factor' at the order of $10\sim 1000$ 
is necessary to account for the CACTUS results. 
The `boost factor' means that the astrophysical
factor calculated by a cored or a cuspy profile in the Sec II
should be enhanced by this factor to give the observed flux. 
The `boost factor' requires a much sharper density profile compared
with the NFW profile, such as a Moore profile \cite{moore}
or a spike profile due to the existence of an intermediate mass
central black hole \cite{blh} in Draco.

In Figs. \ref{10}, \ref{100} and \ref{1000}, we plot the integrated
$\gamma$-ray fluxes above the threshold energy $50 GeV$
within the solid angle $\Delta\Omega=10^{-3}$ as a function 
of the neutralino mass.
The results in Figs. \ref{10}, \ref{100} and \ref{1000}
have enhanced the astrophysical factor by a boost factor of
$10$, $100$ and $1000$ respectively.
Each point in the figure corresponds to a
model with a set of definite SUSY parameters
in the 8-dimensional parameter space which can explain the
CACTUS results constrained by Eqs. (\ref{i50}) and (\ref{i150})
and allowed by all other collider and cosmology constraints.
The scatter of the points represents the uncertainty 
coming from the unknown soft SUSY breaking parameters.
The lines represent the $2\sigma$ constraints of ARGO 
by assuming the final states being the $W^+W^-$, $b\bar{b}$ or
$\tau\bar{\tau}$. For the upper set of line we have
assumed no hadron/photon discrimination at all, while for
the lower set of lines we assume a part of hadrons are rejected
based on neural network
so that the significance of detection is improved by a quality factor
of $1.6$ \cite{proposal}.
From the figure we can see that if no excess is observed at ARGO 
above 100 GeV a large part of the parameter space
is constrained. 

It is worthwhile commenting on the results here. 
First, if extending the gamma spectrum to lower energies,
we find the CACTUS result is difficult to reconcile with 
the EGRET result \cite{egret} which
did not observe excess at the direction of Draco between 1$\sim$ 10 GeV.
Therefore a hard spectrum is expected to reconcile the EGRET and the CACTUS
results, which requires the dominant annihilation product
be $\tau\bar{\tau}$ \cite{hooper}. 
The hard spectrum leads to more opportunities to
observe the signal in ARGO which can be seen from
the Figs. \ref{10}, \ref{100} and \ref{1000}.
Alternatively one would assume that only about 1 percent of
the present excess is real signal from the annihilation of the dark
matter. In this case we find parameter
space to account for the signal and be consistent with EGRET result
in the range of $250 GeV < m_\chi < 800 GeV$.
The parameter space can be constrained by ARGO only for the $\tau\bar{\tau}$
final states.
Second, the CACTUS result may also imply a monochromatic gamma spectrum at
the energy of about 50 GeV. However, it is found that the branching
ratio that two neutralino annihilate into two photons should be
more than a half to be consistent with the EGRET result,
which is incompatible with the SUSY model \cite{profumo}.
Finally, if we assume that 
only about $\lesssim 1\%$ of the excess
comes from DM annihilation, the signal can be explained without 
introduce any `boost factor' if taking the nonthermal mechanism
into account. This may be a natural assumption, while the confirmation
of the gamma events from DM annihilation requires an instrument with
better angular resolution, such as GLAST \cite{profumo} to suppress
the background.

\section{Summary and Conclusion}

In this paper we have discussed the possibility of constraining the signal 
observed by the CACTUS experiment at the ground based EAS detector, 
ARGO. We assume the excess of gamma rays observed at CACTUS is 
produced by supersymmetric dark matter annihilation.
We then explore the SUSY parameter space to give signal
consistent with the CACTUS result and discuss the possibility
to constrain the parameter space at ARGO. Our calculation shows
that, depending on the gamma spectrum, 
ARGO will be able to constrain a large part of the parameter space
if no signal is detected for one year observation. 

If the CACTUS signal is finally confirmed, the implication on dark matter
is dramatic. The central cusp of the dark halo at Draco should 
be much sharper than that of a NFW profile. The neutralino mass 
should be at the range of $100\sim 400$ GeV
to explain the signal of CACTUS. Furthermore, the spectrum of
the annihilation gamma ray should be very hard in order to be consistent
with the EGRET null result at the direction of Draco at the energy
range between 1 GeV and 10 GeV.

\begin{acknowledgments}
We thank Xuelei Chen for helpful discussions.
This work is supported in part by the NSF of China under the grant
Nos. 10575111, 10105004, 10120130794, 90303004.
\end{acknowledgments}


\begin{thebibliography}{999}

\bibitem{review}
  G.~Jungman, M.~Kamionkowski and K.~Griest,
  Phys.\ Rept.\  {\bf 267} (1996) 195;
  G.~Bertone, D.~Hooper and J.~Silk,
  Phys.\ Rept.\  {\bf 405}, 279 (2005).

\bibitem{berg}
  L.~Bergstr\"om, P.~Ullio and J.~H.~Buckley,
  Astropart.\ Phys.\  {\bf 9}, 137 (1998).

\bibitem{baltz}
 E. A. Baltz, C. Briot, P. Salati, R. Taillet, J. Silk,
Phys.\ Rev.\ D {\bf 61}, 023514 (2000), arXiv:astro-ph/9909112.

\bibitem{pieri}
 N.~Fornengo, L.~Pieri and S.~Scopel,
  Phys.\ Rev.\ D {\bf 70}, 103529 (2004), arXiv:hep-ph/0407342.

\bibitem{evans}
  N.~W.~Evans, F.~Ferrer and S.~Sarkar,
  Phys.\ Rev.\ D {\bf 69}, 123501 (2004), arXiv:astro-ph/0311145.
 
\bibitem{bi}
X.J. Bi, arXiv:astro-ph/0510714.

\bibitem{kou}
S.M. Koushiappas, A.R. Zentner, T. P. Walker,
Phys. Rev. \textbf{D 69}, 043501 (2004).
                                                                                
\bibitem{cactus}
P.~Marleau, TAUP, Zaragoza, Spain, September 2005;  M.~Tripathi,
Cosmic Rays to Colliders 2005, Prague, Czech Republic, September
2005; TeV Particle Astrophysics Workshop, Batavia, USA, July 2005;
M.~Chertok, proceedings of PANIC 05, Santa Fe, USA, October 2005.
                                                                                
\bibitem{young}
L.M.~Young, Astron. J. {\bf 117}, 1758 (1999).

\bibitem{tyler}
  C.~Tyler,
  Phys.\ Rev.\ D {\bf 66} (2002) 023509,  arXiv:astro-ph/0203242.
 
\bibitem{hooper}
L. Bergstr\"om, D. Hooper,
arXiv:hep-ph/0512317.

\bibitem{profumo}
 S. Profumo, M. Kamionkowski,
arXiv:astro-ph/0601249.

\bibitem{glast}
A.Morselli et al., Proc. of the 32nd Rencontres de Moriond (1997).

\bibitem{magic}
C. Baixeras et al., Nucl. Phys. Proc. Suppl., \textbf{114}, 247 (2003).

\bibitem{argo}
A. Aloisio et al.,
Nuovo Cim. 24C, 739 (2001);
G. Di Sciascio et al., [ARGO-YBJ Collaboration],
AIP Conf. Proc. 745, 663 (2005).

\bibitem{pythia}
T. Sj\"ostrand et al., Comput. Phys. Commun. \textbf{135}, 238 (2001).

\bibitem{cepheid}
A. Z. Bonanos, K. Z. Stanek, A.H. Szentgyorgyi, D.D. Sasselov and G.A. Bakos,
Astrophys. J. {\bf 127}, 821 (2004).
                                                                                
\bibitem{nfw}
J.  F.  Navarro, C.  S.  Frenk,  and  S. D.  M.  White,
Mon.  Not. R.   Astron.  Soc. {\bf  275},  56 (1995);  J. F.  Navarro,
C. S. Frenk, and S. D. M. White, Astrophys.  J. {\bf 462}, 563 (1996);
J. F.  Navarro, C. S. Frenk, and  S. D. M. White,  Astrophys.  J. {\bf
490}, 493 (1997).
                                                                                

\bibitem{mash}
S. Mashchenko, H. M. P. Couchman, A. Sills,
Astrophys. J. \textbf{624}, 726 (2005).
S.~Mashchenko, A.~Sills and H.M.P.~Couchman,
arXiv:astro-ph/0511567.
                                                                                
\bibitem{gaisser}
T. Gaisser et al., Proc. of the 27th ICRC (2001).
                                                                                
\bibitem{longair}
M.S. Longair, High Energy Astrophysics, Cambridge
Universeity Press (1992).
                                                                                
\bibitem{gal}
L. Bergstr\"om, P. Ullio, J. Buckley, Astropart. Phys. \textbf{9}, 137 (1998).

\bibitem{cui}
S.W. Cui, H.B. Hu, Proc. of the 29th ICRC (2005);
X.X. Zhou et al. Proc. of the 29th ICRC (2005).

                                                                                
\bibitem{hawc}
 G. Sinnis, A. Smith, J.E. McEnery,
astro-ph/0403096.
                                                                                

\bibitem{fornengo}
N. Fornengo, L. Pieri, S. Scopel,
Phys. Rev. \textbf{D 70}, 103529 (2004).

\bibitem{darksusy}
P. Gondolo, J. Edsjo, P. Ullio, L. Bergstrom, M. Schelke, E.A. Baltz,
JCAP 0407, 008 (2004), astro-ph/0406204.
                                                                                
\bibitem{pdg}
K. Hagiwara et al., Phys. Rev. D \textbf{66}, 010001 (2002).
                                                                                
\bibitem{wmap}
C. L. Bennett \textit{et al.},
Astrophys. J. Suppl. 148, 1 (2003),
arXiv: astro-ph/0302207;
D. N. Spergel \textit{et al.},
Astrophys. J. Suppl. 148, 175 (2003),
arXiv: astro-ph/0302209.
                                                                                

\bibitem{nonther}
R.~Jeannerot, X.~Zhang and R.~H.~Brandenberger,
JHEP {\bf 9912}, 003 (1999);
T.~Moroi and L.~Randall,
Nucl.\ Phys.\ B {\bf 570}, 455 (2000);
W.~B.~Lin, D.~H.~Huang, X.~Zhang and R.~H.~Brandenberger,
Phys.\ Rev.\ Lett.\  {\bf 86}, 954 (2001);
B.~Murakami and J.~D.~Wells,
Phys.\ Rev.\ D {\bf 64}, 015001 (2001);
M.~Fujii and K.~Hamaguchi,
Phys.\ Lett.\ B {\bf 525}, 143 (2002);
G. B. Gelmini, P. Gondolo, 
arXiv: hep-ph/0602230.

\bibitem{moore}
B. Moore,  S. Ghigna,  F. Governato,  G. Lake,  T. Quinn,  J. Stadel,
 P. Tozzi, 1999, ApJ, 524, L19.

\bibitem{blh}
 P. Gondolo, J. Silk,
Phys. Rev. Lett. \textbf{83}, 1719 (1999);
  G.~Bertone, A.~R.~Zentner and J.~Silk,
  Phys.\ Rev.\ D {\bf 72}, 103517 (2005), arXiv:astro-ph/0509565.
  
\bibitem{proposal}
The ARGO-YBJ Project (addendum to the Proposal), ARGO-YBJ collaboration.

\bibitem{egret}
{\tt http://cossc.gsfc.nasa.gov/egret/}






\end{thebibliography}
\end{document}